\documentclass{article}

\usepackage[margin=1.5in]{geometry}
\usepackage[utf8]{inputenc}
\usepackage{natbib}
\usepackage{multirow}
\usepackage{placeins}
\usepackage[colorlinks=True]{hyperref}
\usepackage{graphicx}
\usepackage{subcaption}
\usepackage[T1]{fontenc}
\usepackage[utf8]{inputenc}
\usepackage{url}
\usepackage{amsthm}
\usepackage{float}
\usepackage{color}
\usepackage{amsmath}

\definecolor{darkblue}{rgb}{0.0,0.0,0.55}
\hypersetup{
  colorlinks = true,
  citecolor  = darkblue,
  linkcolor  = darkblue,
  citecolor  = darkblue,
  filecolor  = darkblue,
  urlcolor   = darkblue,
}
\makeatletter
\let\@fnsymbol\@arabic
\makeatother
\title{We Should At Least Be Able To Design Molecules That Dock Well}
\author{Tobiasz Cieplinski\thanks{Jagiellonian University, Poland}, Tomasz Danel\footnotemark[1] , Sabina Podlewska\footnotemark[1], \\Stanisław Jastrzębski\thanks{Molecule.one, Poland}\,\,\footnotemark[1]}

\date{}

\begin{document}

\maketitle

\begin{abstract}
Designing compounds with desired properties is a key element of the drug discovery process. However, measuring progress in the field has been challenging due to lack of realistic retrospective benchmarks, and the large cost of prospective validation. To close this gap, we propose a benchmark based on docking, a popular computational method for assessing molecule binding to a protein. Concretely, the goal is to generate drug-like molecules that are scored highly by SMINA, a popular docking software. We observe that popular graph-based generative models fail to generate molecules with a high docking score when trained using a realistically sized training set. This suggest a limitation of the current incarnation of models for de novo drug design. Finally, we also include simpler tasks in the benchmark based on a simpler scoring function. We release the benchmark as an easy to use package available at \href{https://github.com/cieplinski-tobiasz/smina-docking-benchmark}{https://github.com/cieplinski-tobiasz/smina-docking-benchmark}. We hope that our benchmark will serve as a stepping stone towards the goal of automatically generating promising drug candidates.
\end{abstract}

\section{Introduction}

Designing compounds with some desired chemical properties is the central challenge in the drug discovery process~\citep{sliwoski2014,elton2019}. De novo drug design is one of the most successful computational approach that involves generating new potential ligands \emph{from scratch}, which avoids enumerating explicitly the vast space of possible structures. Recently, deep learning has unlocked new progress in drug design. Promising results using deep generative models have been shown in generating soluble~\citep{gvae}, bioactive~\citep{segler_rnn}, and drug-like~\citep{jin2018junction} molecules.

A key challenge in the field of drug design is the lack of realistic benchmarks~\citep{elton2019}. Ideally, the generated molecule by a de novo method should be tested in the wet lab for the desired property. In practice, typically, a proxy is used. For example, the octanol-water partition coefficient or bioactivity is predicted using a computational model~\citep{segler_rnn,gvae}. However, these models are often too simplistic~\citep{elton2019}. This is aptly summarized in \citet{coley2019} as ``The current evaluations for generative models do not reflect the complexity of real discovery problems.''. In contrast to drug design, more realistic benchmarks have been used in the design of photovoltaics~\citep{pyzer2015} or in the design oof molecules with certain excitation energies~\citep{Sumita2018}, where a physical calculation was carried out to both train models, and to evaluate generated compounds. 

Our main contribution is a realistic benchmark for de novo drug design. We base our benchmark on docking, a popular computational method for predicting molecule binding to a protein. Concretely, the goal is to generate molecules that are scored highly by SMINA~\citep{smina}. We picked \citet{smina} due to its popularity and being available under a free license. While we focus on de novo drug design, our methodology can be extended to evaluate retrospectively other approaches to designing molecules. Code to reproduce results and evaluate new models is available online at \href{https://github.com/cieplinski-tobiasz/smina-docking-benchmark}{https://github.com/cieplinski-tobiasz/smina-docking-benchmark}. 

Our second contribution is exposing limitation of currently popular de novo drug design methods for generating bioactive molecules. When trained using a few thousands compounds, a typical training set size, the tested methods fail to generate highly active structures according to the docking software. The highest scoring molecules in most cases did not outperform top 10\% molecules found in either the ZINC database or the training set. This suggest we should exercise caution when applying them in drug discovery pipelines, where we seldom have larger number of known ligands. We hope our benchmark will serve as a stepping stone to further improve these promising models.

The paper is organised as follows. We first discuss prior work and introduce our benchmark. Next, we use our benchmark to evaluate two popular models for de novo drug design. Finally, we analyse why the tested models fail on the most difficult version of the benchmark.

\section{Docking-based benchmark}

\begin{figure}
    \centering
    \includegraphics[width=\textwidth]{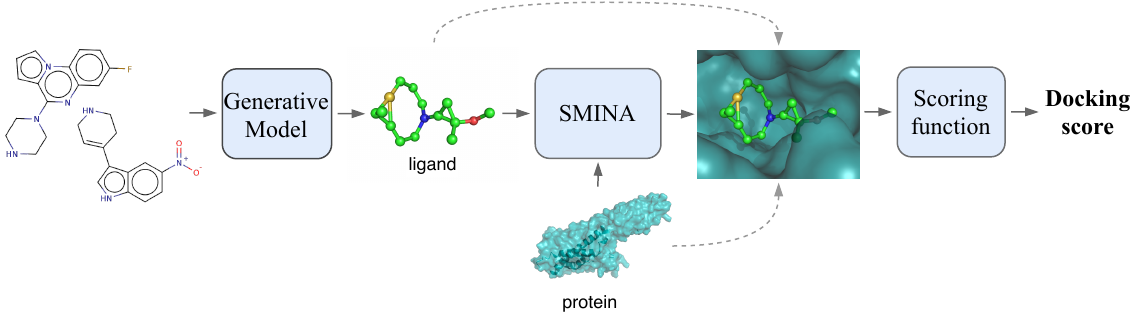}
    \caption{Visualization of the proposed docking-based benchmark for de-novo drug design methods. First, the proposed molecule (leftmost) is docked to the target's binding site using SMINA, a popular docking software. In the most difficult version of the benchmark, the final score is computed based on the ligand pose using the default SMINA docking score. }
    \label{fig:benchmark_visualization}
\end{figure}

We begin by briefly discussing prior work and motivation. Next, we introduce our benchmark.

\subsection{Why do we need yet another benchmark?}

Standardized benchmarks are critical to measure progress in any field. Development of large-scale benchmarks such as the ImageNet was critical for the recent developments in artificial intelligence~\citep{imagenet_cvpr09,wang2018}. Many new methods for de novo drug design are conceived every year, which motivates the need for a systematic and efficient way to compare them~\citep{Schneider2019}.

De novo drug design methods are typically evaluated using \emph{proxy tasks} that circumvent the need to test the generated compounds experimentally~\citep{jin2018junction, you2018, maziarka2020,gvae,bombarelli2016}. Optimizing the octanol-water partition coefficient (logP) is a common example. The logP coefficient is commonly computed using an atom-based method that involves summing contribution of individual atoms~\citep{wildman1999,jin2018junction}, which is available in the RDKit package~\citep{rdkit}. Due to the fact that it is easy to optimize the atom-based method by producing unrealistic molecules~\citep{guacamol}, a version that heuristically penalizes hard to synthesize compounds is used in practice~\citep{jin2018junction}. This example illustrates the need to develop more realistic ways to benchmark these methods. Another example is QED score~\citep{bickerton2012} which is designed to capture \emph{druglikeliness} of a compound. Finally, some approacches use a model (e.g. a neural network) to predict bioactivity of the generated compounds~\cite{segler_rnn}. Similarly to logP, these two tasks are also possible to optimize while producing unrealistic molecules. This is aptly summarized in \citet{coley2019} as 

\begin{quote}
    ``The current evaluations for generative models do not reflect the complexity of real discovery problems.''
\end{quote}

Interestingly, besides the aforementioned proxy tasks, more realistic proxy tasks are rarely used in the context of evaluating de novo drug design methods. This is in contrast to evaluation of generative models for generating photovoltaics~\citep{pyzer2015} or molecules with certain excitation energies~\citep{Sumita2018}. One notable exception is \citet{armstrong2018} who try to generate compounds that are active according to the DrugScore~\citep{gerd2011}, and then evaluate the generated compounds using rDock~\citep{RuizCarmona2014}. This lack of the overall diversity and realism in the typically used evaluation methods motivates us to propose our benchmark. 

\subsection{Docking-based benchmark}

Our docking-based benchmark is defined by: (1) docking software that computes for a generated compound its pose in the binding site, (2) a function that scores the pose, (3) a training set of compounds with an already computed docking score. 

The goal is to generate a given number of molecules that achieve the maximum possible docking score. For the sake of simplicity, we do not impose limits on the distance of the proposed compounds to training set. Thus a simple baseline is to return the training set. Finding similar compounds that have a higher docking score is already prohibitively challenging for current state-of-the-art methods. As the field progresses, our benchmark can be easily extended to account for the similarity between the generated compounds and the training set.

Finally, we would like to stress that the benchmark is not limited to de novo methods. The benchmark is applicable to any other approaches such as virtual screening. The only limitation required for a fair comparison is that docking is performed only on the supplied training set.

\subsection{Instantiation}
\label{sec:instatiation}

As a concrete instantiation of our docking-based benchmark, we use SMINA v. 2017.11.9~\citep{smina} due to its wide-spread use and being offered under a free license. To create the training set, we download from the ChEMBL~\citep{chembl2016} database molecules tested against popular drug-targets: 5-HT1B, 5-HT2B, ACM2, and CYP2D6. For 5-HT1B the final dataset consists in $1891$ molecules, out of which $1148$ are active (Ki < 100nm) and $743$ are inactive molecules (Ki > 1000nm). We list sizes of the datasets in Table~\ref{tab:dataset_sizes}.

We dock each molecule using default settings in SMINA to a manually selected binding site coordinate. Protein structures were downloaded from the Protein Database Bank, cleaned and prepared for docking using Schrödinger modeling package. The resulting protein structures are provided in our code repository. We describe further details on the preparation of the datasets in Appendix~\ref{app:dataset_details}.

\begin{table}[]
\centering
\begin{tabular}{ |c|c|c|c|c| } 
 \hline
 & 5HT1B & 5HT2B & ACM2 & CYP2D6 \\
 \hline
 Dataset size & 1878 & 1193 & 2337 & 4199 \\
 \hline
 \# Actives & 1139 & 656 & 1300 & 343 \\
 \hline
 \# Inactives & 739 & 537 & 1037 & 3856 \\
 \hline
\end{tabular}
\caption{Sizes of the dataset used in the benchmark. The corresponding test dataset comprises of $10\%$ of the whole dataset, and the rest of it is used in training.}
 \label{tab:dataset_sizes}
\end{table}

Starting from the above, we define the following three variants of the benchmark. In the first variant, the goal is to propose molecules that achieve the smallest SMINA docking score used in score only mode, defined as follows:

\begin{equation*}
\label{eq:docking_score}
    \begin{split}
    \textnormal{Docking score} =& -0.035579 \cdot \textnormal{gauss}(o=0, w=0.5) \\
     & -0.005156 \cdot \textnormal{gauss}(o=3, w=2) \\
     & +0.840245 \cdot \textnormal{repulsion} \\
     & -0.035069 \cdot \textnormal{hydrophobic} \\
     & -0.587439 \cdot \textnormal{non\_dir\_h\_bond}
    \end{split},
\end{equation*}

where all terms are computed based on the final docking pose. The first three terms measure the steric interaction between ligand and the protein. The fourth and the fifth term look for hydrophobic and hydrogen bonds between the ligand and the protein. We include in Appendix~\ref{app:smina_scoring_function} a detailed description of all the terms.

Next, we propose two simpler variants of the benchmark based on individual terms in the SMINA scoring function. In the \emph{Repulsion} task, the goal is to only minimize the repulsion component, which is defined as:

$$
\text{repulsion}(a_1, a_2) =
\begin{cases}
    d_\text{diff}(a_1, a_2)^2, & d_\text{diff}(a_1, a_2) < 0 \\
    0, & \text{otherwise}
\end{cases}
$$

where $d_\text{diff}(a_1, a_2)$ is the distance between the atoms minus the sum of their van der Waals radii. Distance unit is Angstrom ($10^{-10}\text{m}$).

The third task, \emph{Hydrogen Bond Task}, is to maximize the non\_dir\_h\_bond term:

$$
\text{non\_dir\_h\_bond}(a_1, a_2) =
\begin{cases}
    0, & (a_1, a_2) \,\text{do not form hydrogen bond} \\
    1, & d_\text{diff}(a_1, a_2) < -0.7 \\
    0, & d_\text{diff}(a_1, a_2) \geq 0 \\
    \frac{d_\text{diff}(a_1, a_2)}{-0.7} & \text{otherwise}.
\end{cases}
$$

To make the results more stable, we average the score over the top 5 best-scoring binding poses. Finally, to make the benchmark more realistic, we filter the generated compounds using the Lipinski rule, and discard molecules with molecular weight lower than 100.

\subsection{Novelty criterion}

To avoid trivial solutions, we filter out generated compounds that are similar to the training set. To set the similarity threshold, we downloaded $10^6$ compounds from the ZINC~\citep{irwin2005ZINC} database and calculated their similarity to each of the training sets used in our benchmark. To evaluate similarity we used the ECFP~\citep{Rogers2010ECFP} representation with the radius set to $2$ and the number of bits of $1024$. Based on the results, we simply set the similarity threshold to $0.2$, which is approximately the $5$th percentile of the similarity of ZINC molecules to each of the training sets.

\subsection{ZINC baseline}

The premise behind de novo drug discovery is that it enables access to structurally novel and potent molecules. To contextualize results in the benchmark, we included as the baseline sampling from the subset of ZINC database containing 9,204,719 molecules\footnote{Downloaded from \href{http://ZINC15.docking.org}{http://ZINC15.docking.org}.}. We selected molecules having following properties: 3D representation, standard reactivity, in-stock purchasability, ref pH, charges from -2 to 2 inclusive and used drug-like preferred subset. For each protein we have sampled a set of molecules from aforementioned ZINC subset of protein's training set size. In each task we compare to the mean score, the top 10\%, and the top 1\% of scores.

\subsection{Diversity}

\begin{figure}
    \centering
     \begin{subfigure}[b]{0.45\textwidth}
         \centering
         \includegraphics[width=\textwidth]{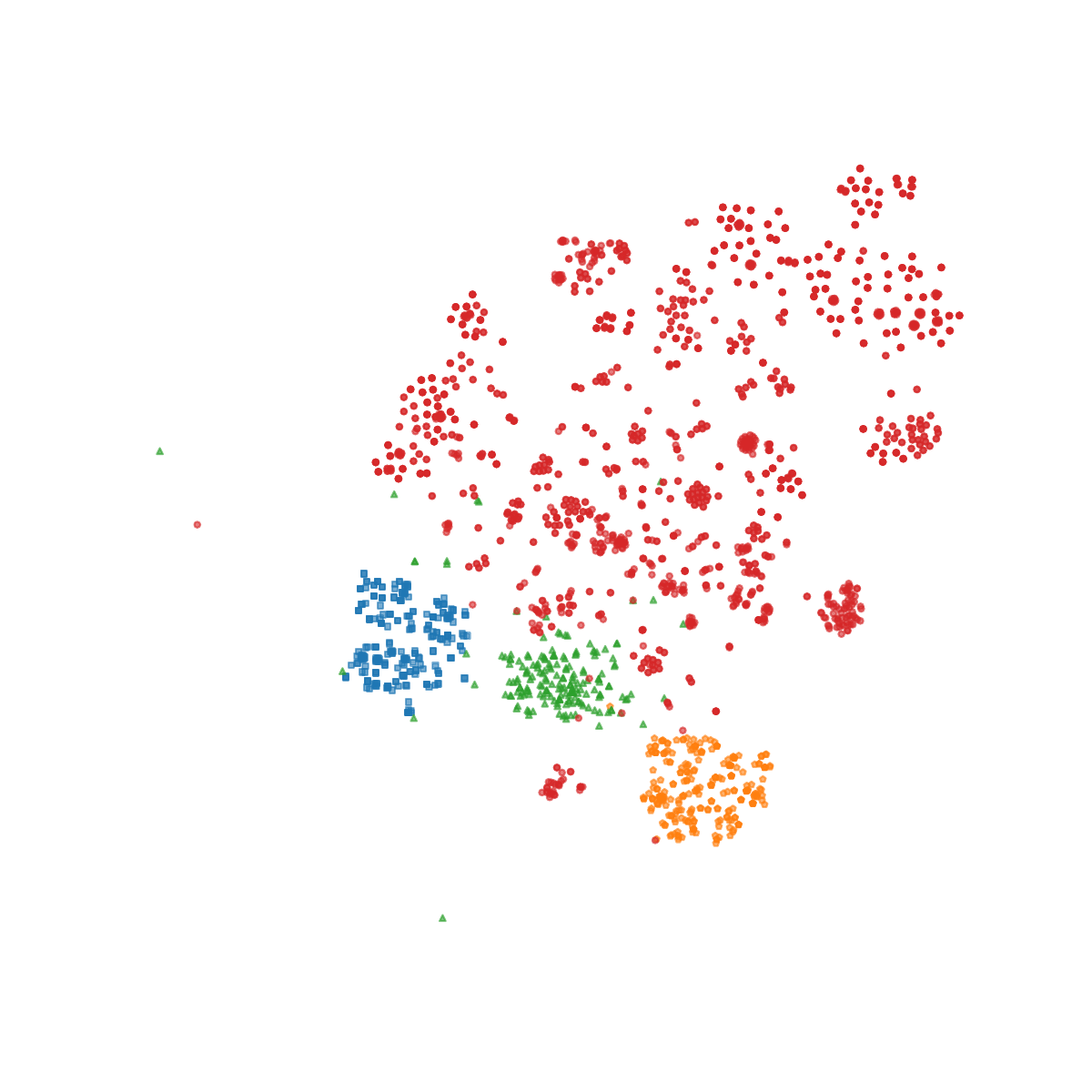}
         \caption{5HT1B}
     \end{subfigure}
     \hfill
     \begin{subfigure}[b]{0.45\textwidth}
         \centering
         \includegraphics[width=\textwidth]{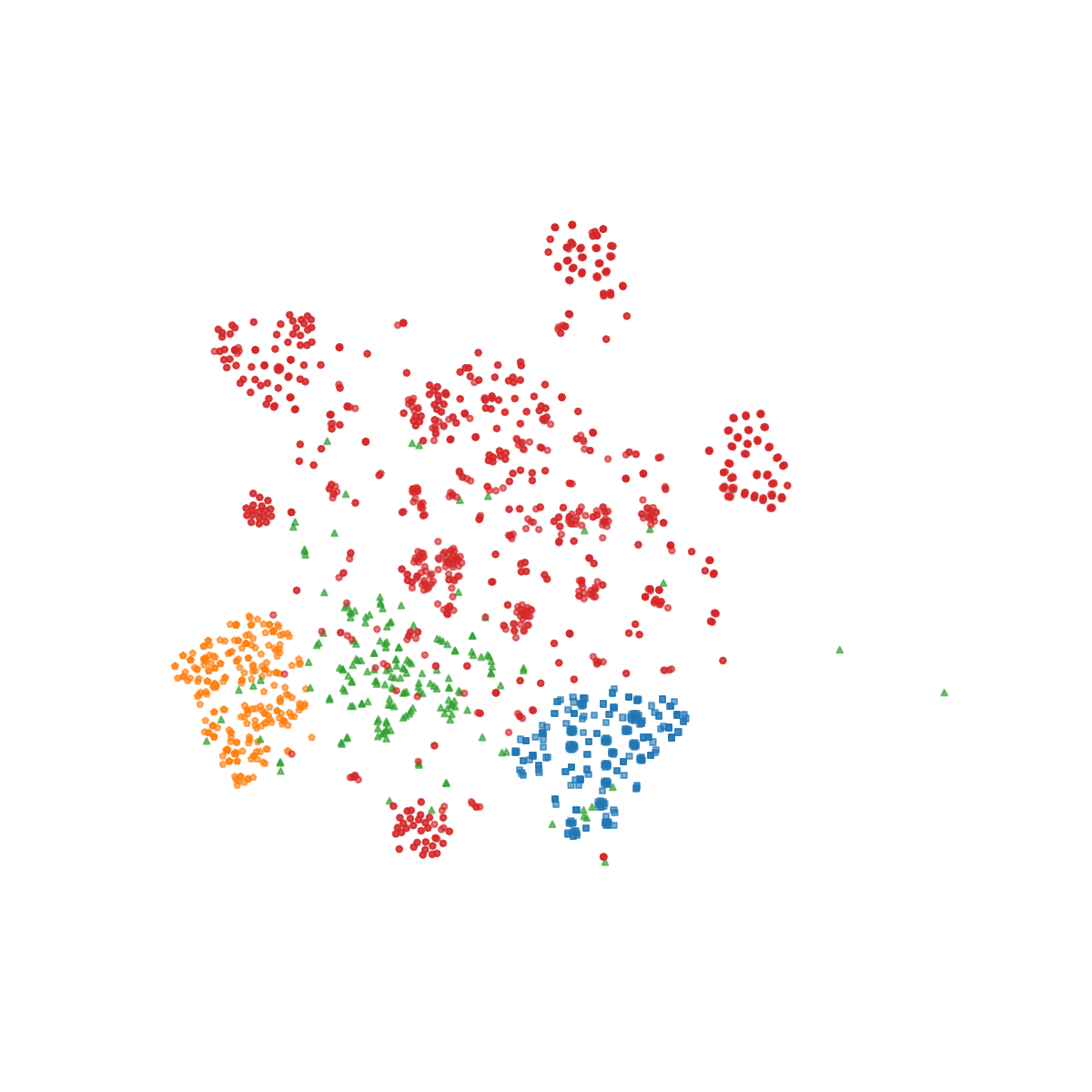}
         \caption{5HT2B}
     \end{subfigure}
     \hfill
     \begin{subfigure}[b]{0.45\textwidth}
         \centering
         \includegraphics[width=\textwidth]{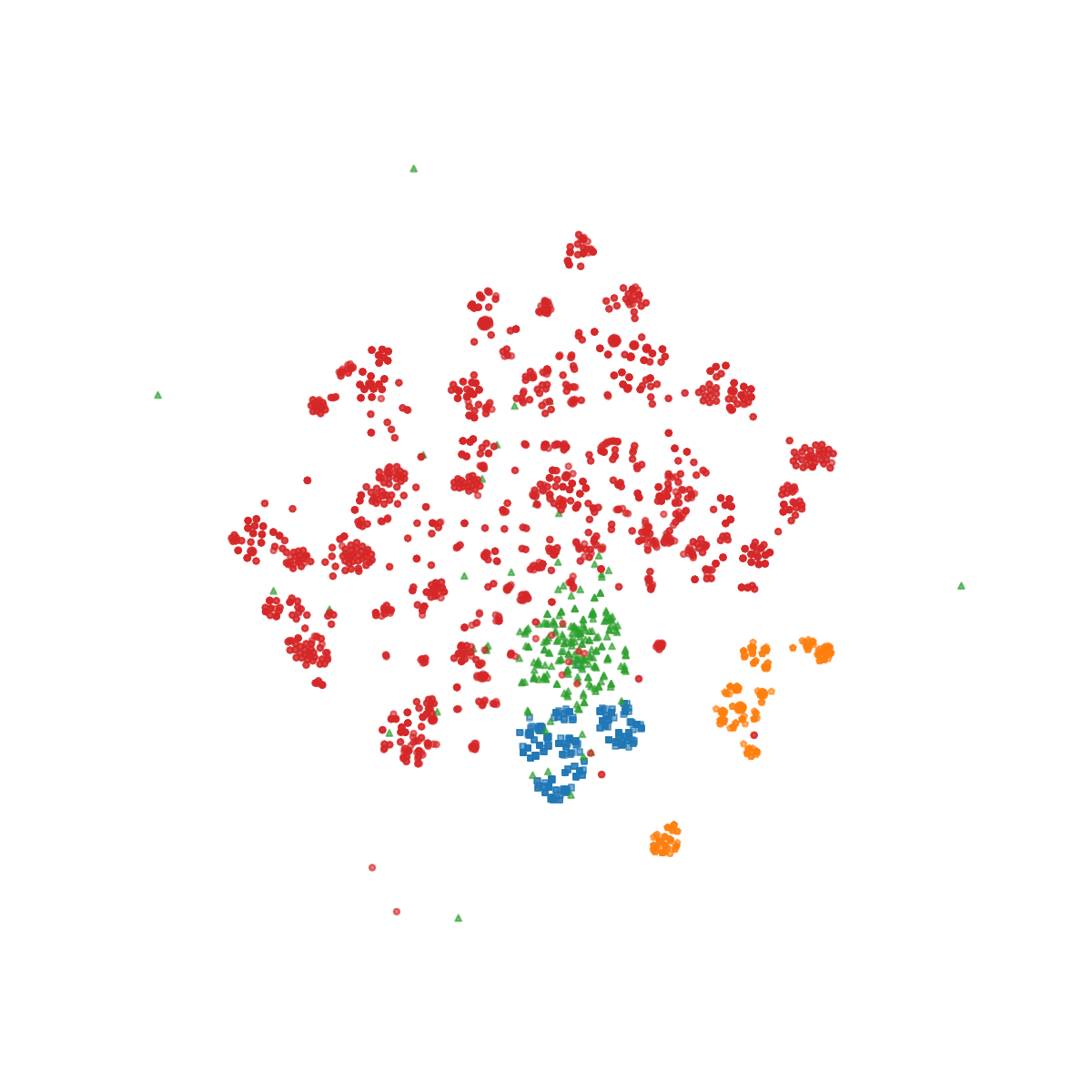}
         \caption{ACM2}
     \end{subfigure}
     \hfill
     \begin{subfigure}[b]{0.45\textwidth}
         \centering
         \includegraphics[width=\textwidth]{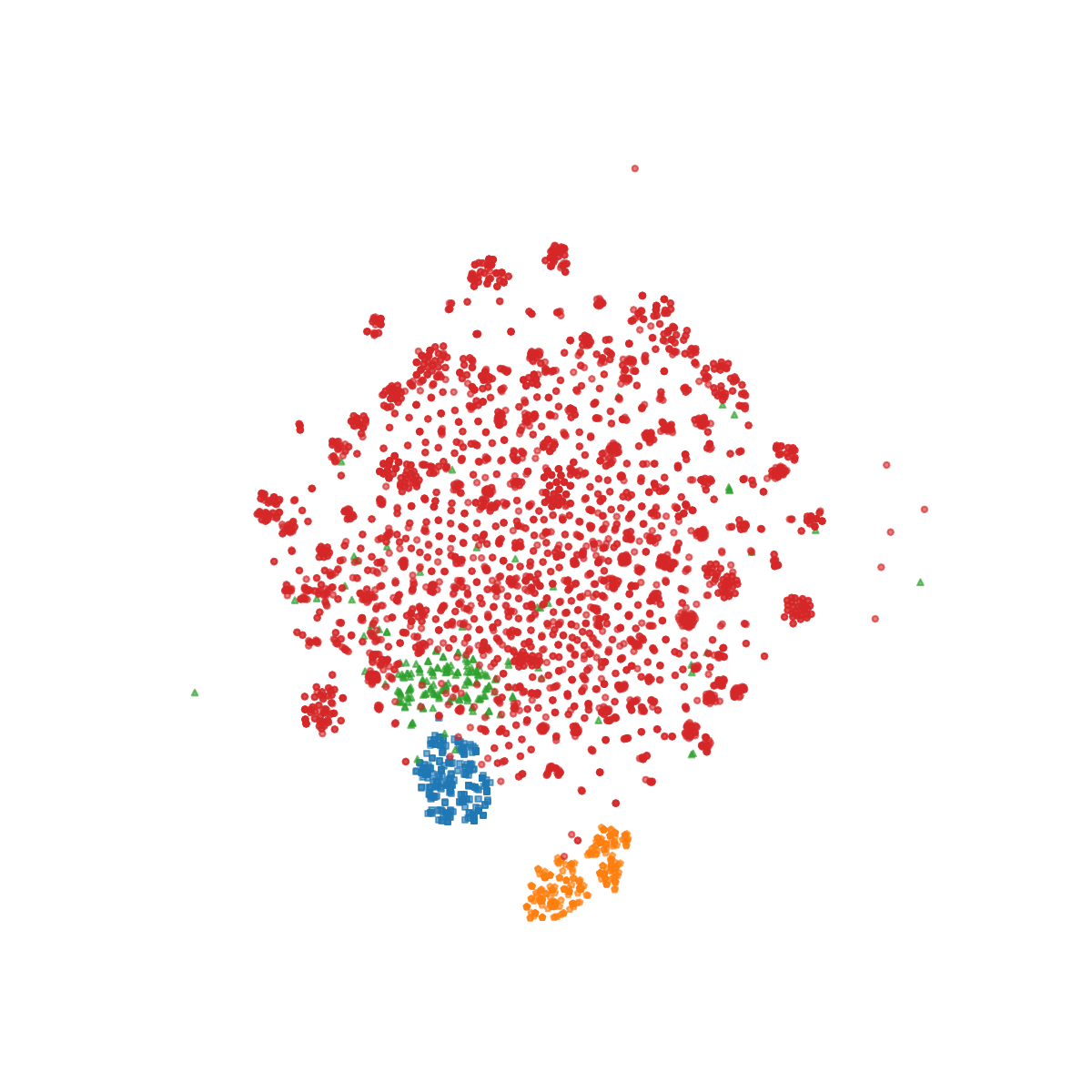}
         \caption{CYP2D6}
     \end{subfigure}
    \caption{t-SNE maps of compound fingerprints (ECFP) for each protein. The training set is marked with red dots, and the compounds generated by REINVENT by optimizing different targets are colored in blue (\textsc{Docking Score Function}), orange (\textsc{Hydrogen Bonding}), and green (\textsc{Repulsion}).}
    \label{fig:tsne}
\end{figure}

To better understand the performance of each model, besides the mean score, we also evaluate the diversity of the proposed molecules. Concretely, we compute the mean Tanimoto distance between all pairs of molecules in the generated sample. We use the 1024-bit ECFP representation~\citep{Rogers2010ECFP} with radius 2. The score is reported in the benchmark along with the docking score results. We observe that the optimized models narrow down to a less diverse subspace of compounds that are dissimilar to the training set. This can also be observed in the t-SNE plots of the generated compounds compared to the training set (Figure~\ref{fig:tsne}). The small focused clouds of compounds generated using different optimization targets always concentrate at one side of the map. This suggests that there is similar bias of the model independent of the optimization target, which can be the ChEMBL prior of the REINVENT model. Besides that observation, we note that the generated compounds are less diverse, showing similarity only inside one generated batch (the same optimization target).

\subsection{When is a task solved?}

In the experiments, we compare to two baselines: (i) random compounds from ZINC as the baseline, and (ii) compounds from the training set. In each case we report the mean score, the top 10\% of scores, and the top 1\% of scores. We also report diversity of the results. 

Roughly speaking, \textit{we consider a given task solved if the generated molecules exceed the top 1\% score found in the ZINC database, while achieving at least the same diversity as observed in the training set.} This criterion is necessarily arbitrary. It is inspired by a natural baseline -- sampling several thousands of compounds from the ZINC database.

\section{Results and discussion}

In this section, we evaluate three popular models for de novo drug design on our docking-based benchmark.

\subsection{Models}

We compare three popular models for de novo drug design. Chemical Variational Autoencoder (CVAE)~\citep{cvae} applies Variational Autoencoder~\citep{vae} by representing molecules as strings of characters (using SMILES encoding). This approach was later extended by Grammar Variational Autoencoder (GVAE)~\citep{gvae}, which ensures that generated compounds are grammatically correct. The third model, REINVENT~\citep{olivecrona2017molecular}, is a recurrent neural network trained using reinforcement learning for molecular optimization.

\subsection{Experimental details}

To generate active compounds, we follow a similar approach to one in~\citet{jin2018junction}, disregarding the penalty for insufficient similarity. Analogous methods using sparse Gaussian Process instead of multilayer perceptron are also employed in \citet{cvae} and \citet{gvae}. First, we fine-tune a given generative model for $5$ epochs on the training set ligands, starting from weights made available by the authors\footnote{Available at \href{https://github.com/aspuru-guzik-group/chemical_vae/tree/master/models/ZINC}{https://github.com/aspuru-guzik-group/chemical\_vae/tree/master/models/ZINC} and at \href{https://github.com/mkusner/grammarVAE/tree/master/pretrained}{https://github.com/mkusner/grammarVAE/tree/master/pretrained}.}. All hyperparameters are set to default values used in \citet{cvae} and \citet{gvae}. Additionally, we use the provided scores to train a multilayer perceptron (MLP) to predict the target (e.g. the SMINA scoring function) based on the latent space representation of the molecule. 

For CVAE and GVAE, to generate compounds, we first take a random sample from the latent space by sampling from a Gaussian distribution with the standard deviation of $1$ and the mean of $0$. Starting from this point in the latent space, we take 50 gradient steps to optimize the output of the MLP. Based on this approach we generate 250 compounds from the model.

For the REINVENT model, we use pretrained weights on the ChEMBL database provided by \citet{olivecrona2017molecular}. As there is no latent space in this model, we train a random forest model to predict the target directly from the molecule structure. We use the ECFP fingerprint to encode the molecule~\citep{Rogers2010ECFP}. The reward is computed based on the random forest prediction multiplied by the QED score calculated using RDKit.

All other experimental details, including hyperparameter values used in the experiments, can be found in Appendix~\ref{app:hyperparameters}.

\begin{table}[H]

\begin{subtable}[h]{0.95\textwidth}
\centering
\begin{tabular}{ |p{2.0cm}|p{1.1cm}p{1.1cm}|p{1.1cm}p{1.1cm}|p{1.1cm}p{1.1cm}|p{1.1cm}p{1.1cm}|} 
 \hline
 & \multicolumn{2}{|l|}{5HT1B} & \multicolumn{2}{|l|}{5HT2B} & \multicolumn{2}{|l|}{ACM2} & \multicolumn{2}{|l|}{CYP2D6} \\
 \hline
 CVAE & -4.647 & (0.907) & -4.188 & (0.913) & -4.836 & (0.905) & - & - \\
 GVAE & -4.955 & (0.901) & -4.641 & (0.887) & -5.422 & (0.898) & - & - \\
 REINVENT & \textbf{-9.774} & (0.506) & \textbf{-8.657} & (0.455) & \textbf{-9.775} & (0.467) & \textbf{-8.759} & (0.626) \\
 \hline
Train (50\%) & -8.541 & (0.850) & -7.709 & (0.878) & -6.983 & (0.868) & -6.492 & (0.897) \\
Train (10\%) & -10.837 & (0.749) & -9.769 & (0.831) & -8.976 & (0.812) & -9.256 & (0.869) \\
Train (1\%) & -11.493 & (0.859) & -10.023 & (0.746) & -10.003 & (0.773) & -10.131 & (0.763) \\

\hline
ZINC (50\%) & -7.886 & (0.884) & -7.350 & (0.879) & -6.793 & (0.873) & -6.240 & (0.883) \\
ZINC (10\%) & -9.894 & (0.862) & -9.228 & (0.851) & -8.282 & (0.860) & -8.787 & (0.853) \\
ZINC (1\%) & -10.496 & (0.861) & -9.833 & (0.838) & -8.802 & (0.840) & -9.291 & (0.894) \\

\hline
\end{tabular}
\caption{\textsc{Docking Score Function} ($\downarrow$)}
 \label{tab:real_docking_benchmark_results}
\end{subtable}

\begin{subtable}[h]{0.95\textwidth}
\centering
\begin{tabular}{ |p{2.0cm}|p{1.1cm}p{1.1cm}|p{1.1cm}p{1.1cm}|p{1.1cm}p{1.1cm}|p{1.1cm}p{1.1cm}|} 
 \hline
 & \multicolumn{2}{|l|}{5HT1B} & \multicolumn{2}{|l|}{5HT2B} & \multicolumn{2}{|l|}{ACM2} & \multicolumn{2}{|l|}{CYP2D6} \\
 \hline
 CVAE & \textbf{1.148} & (0.919) & \textbf{1.001} & (0.914) & \textbf{1.132} & (0.908) & \textbf{2.234} & (0.914) \\
 GVAE & 1.361 & (0.910) & 1.159 & (0.942) & 1.383 & (0.917) & - & - \\
 REINVENT & 1.544 & (0.811) & 1.874 & (0.859) & 2.262 & (0.845) & 2.993 & (0.858) \\
 \hline
Train (50\%) & 2.099 & (0.845) & 1.792 & (0.881) & 1.434 & (0.863) & 6.508 & (0.895) \\
Train (10\%) & 0.835 & (0.863) & 0.902 & (0.893) & 0.779 & (0.888) & 2.823 & (0.904) \\
Train (1\%) & 0.550 & (0.858) & 0.621 & (0.963) & 0.553 & (0.921) & 1.284 & (0.956) \\
\hline
ZINC (50\%) & 1.803 & (0.878) & 1.677 & (0.882) & 1.665 & (0.879) & 5.786 & (0.880) \\
ZINC (10\%) & 0.840 & (0.880) & 0.865 & (0.896) & 0.792 & (0.881) & 2.348 & (0.887) \\
ZINC (1\%) & 0.613 & (0.941) & 0.625 & (0.922) & 0.612 & (0.938) & 1.821 & (0.880) \\
\hline
\end{tabular}
\caption{\textsc{Repulsion} ($\downarrow$)}
\end{subtable}

\begin{subtable}[h]{0.95\textwidth}
\centering
\begin{tabular}{ |p{2.0cm}|p{1.1cm}p{1.1cm}|p{1.1cm}p{1.1cm}|p{1.1cm}p{1.1cm}|p{1.1cm}p{1.1cm}|} 
 \hline
 & \multicolumn{2}{|l|}{5HT1B} & \multicolumn{2}{|l|}{5HT2B} & \multicolumn{2}{|l|}{ACM2} & \multicolumn{2}{|l|}{CYP2D6} \\
 \hline
 CVAE & 1.089 & (0.915) & 1.168 & (0.909) & 0.881 & (0.907) & 0.539 & (0.908) \\
 GVAE & \textbf{4.152} & (0.921) & \textbf{2.954} & (0.912) & 2.567 & (0.927) & \textbf{2.732} & (0.902) \\
  REINVENT & 3.795 & (0.626) & 2.451 & (0.580) & \textbf{3.520} & (0.480) & 1.304 & (0.574) \\
 \hline
Train (50\%) & 1.069 & (0.843) & 0.668 & (0.882) & 0.296 & (0.871) & 0.684 & (0.892) \\
Train (10\%) & 2.934 & (0.751) & 2.327 & (0.816) & 1.444 & (0.896) & 2.061 & (0.884) \\
Train (1\%) & 3.351 & (0.825) & 3.586 & (0.575) & 2.519 & (0.852) & 2.700 & (0.917) \\
\hline
ZINC (50\%) & 1.114 & (0.879) & 0.871 & (0.882) & 0.512 & (0.877) & 0.660 & (0.877) \\
ZINC (10\%) & 3.623 & (0.873) & 2.674 & (0.887) & 2.449 & (0.874) & 1.831 & (0.878) \\
ZINC (1\%) & 5.743 & (0.928) & 3.545 & (0.935) & 3.253 & (0.940) & 2.115 & (0.861) \\
\hline
\end{tabular}
\caption{\textsc{Hydrogen Bonding} ($\uparrow$)}
\end{subtable}

 \caption{Results on the three molecule generation tasks, each rerun for four different proteins, composing our docking-based benchmark. The key task is \textsc{Docking Score Function} in which the goal is to optimize docking score against a given protein target. Each cell reports the mean score for $250$ generated molecules in each task. In the parenthesis, the internal diversity of generated molecules is reported (see text for details). The tested models tend to improve upon the mean score in the ZINC database (ZINC). However, they generally do not improve upon the top molecules from ZINC; ZINC ($10\%$) and ZINC ($1\%$) show the top $10\%$ of scores and the top $1\%$ of scores. Missing results ("-") indicate that the model failed to generate $250$ molecules that satisfy drug-like filters (see text for details).}
  \label{tab:docking_benchmark_results}
\end{table}

\subsection{Results}
\label{sec:results}

Table~\ref{tab:docking_benchmark_results} summarizes the results on all three tasks. Recall that we generally consider a given task solved if the generated molecules exceed the top 1\% score found in the ZINC database, while achieving at least the same diversity as in the training set. Below we make several observations.

\paragraph{Docking Score Function task} The key task in the benchmark is \textsc{Docking Score Function}. We observe that CVAE and GVAE models fail to generate compounds that achieve higher docking score compared to the mean docking score in the ZINC dataset ($-8.785$ for 5-HT1B compared to $-4.647$ and $-4.955$ achieved by CVAE and GVAE, respectively). The REINVENT model achieves much better performance ($-9.774$ for 5-HT1B). However, while docking scores attained by the molecules generated by REINVENT generally outperform the mean docking  in the ZINC dataset and the training set, they fall short of outperforming the top 10\% molecules found in ZINC ($-9.894$ for 5-HT1B, with the exception of ACM2). We also draw attention to the fact that the generated molecules by REINVENT are markedly less diverse than the diversity of the training set ($0.506$ mean Tanimoto distance compared to $0.787$ in the training set).

These results suggest that generative models applied to de novo drug discovery might require substantial more data to generate well-binding compounds than is typically available for training. In the key \textsc{Docking Score Function} task, models generally fail to outperform the top $10\%$ from the ZINC database. It should worry us that optimizing for docking score, which seems to be a simpler target than true biological binding affinity, is already challenging given realistically sized training sets (between $1193$ and $4199$ molecules).

\paragraph{Repulsion task} Interestingly, REINVENT performs significantly worse than GVAE and CVAE on the \textsc{Repulsion} task. All models fail to outpeform top $10\%$ found in the ZINC dataset. We observe markedly lower diversity of molecules generated by REINVENT compared to the training set.

\paragraph{Hydrogen Bonding task} The \textsc{Hydrogen Bonding} task is the simplest, and both GVAE and REINVENT generate molecules that almost match the top $1\%$ molecules found in the ZINC database and the training set. We again observe relatively low diversity of molecules generated by REINVENT. 

\begin{figure}[H]
    \centering
    \includegraphics[width=\textwidth]{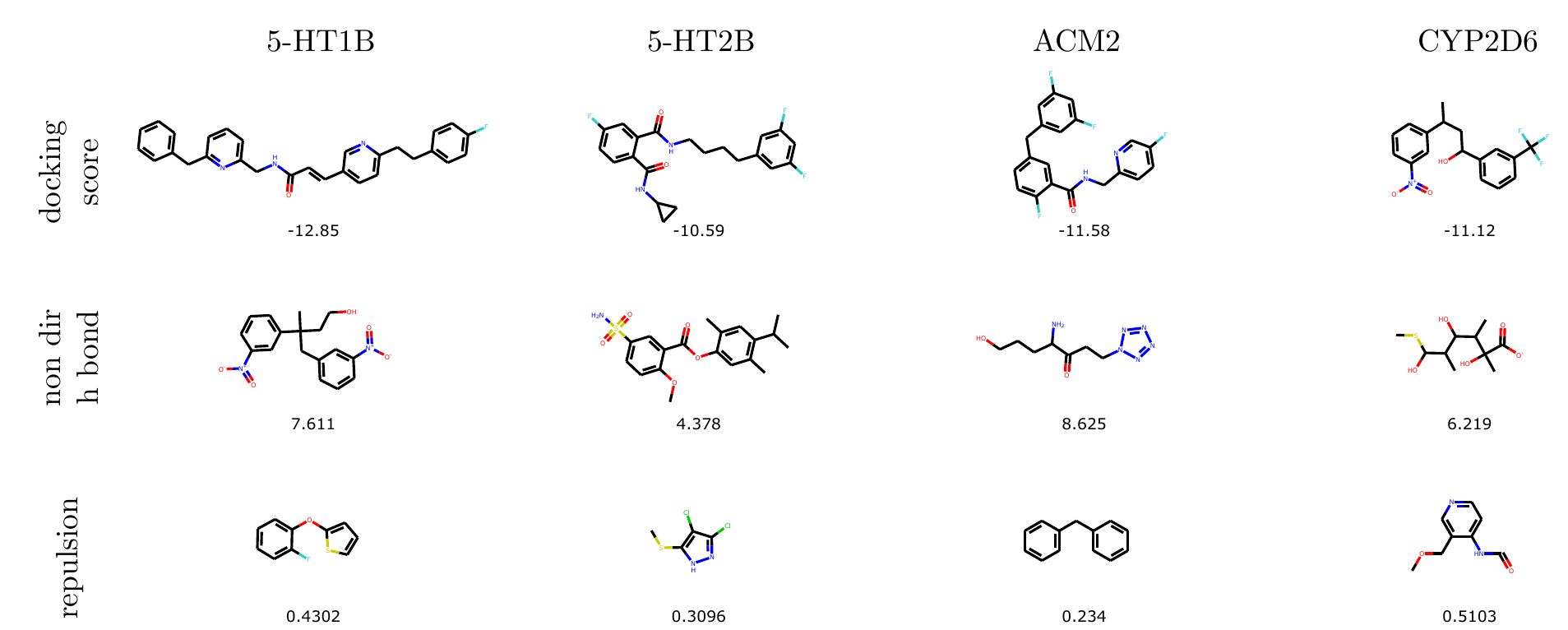}
    \caption{Best scoring molecules generated by REINVENT in each of the three tasks composing the benchmark. }
    \label{fig:benchmark_visualization}
\end{figure}

\begin{figure}
    \centering
    \includegraphics[width=\textwidth]{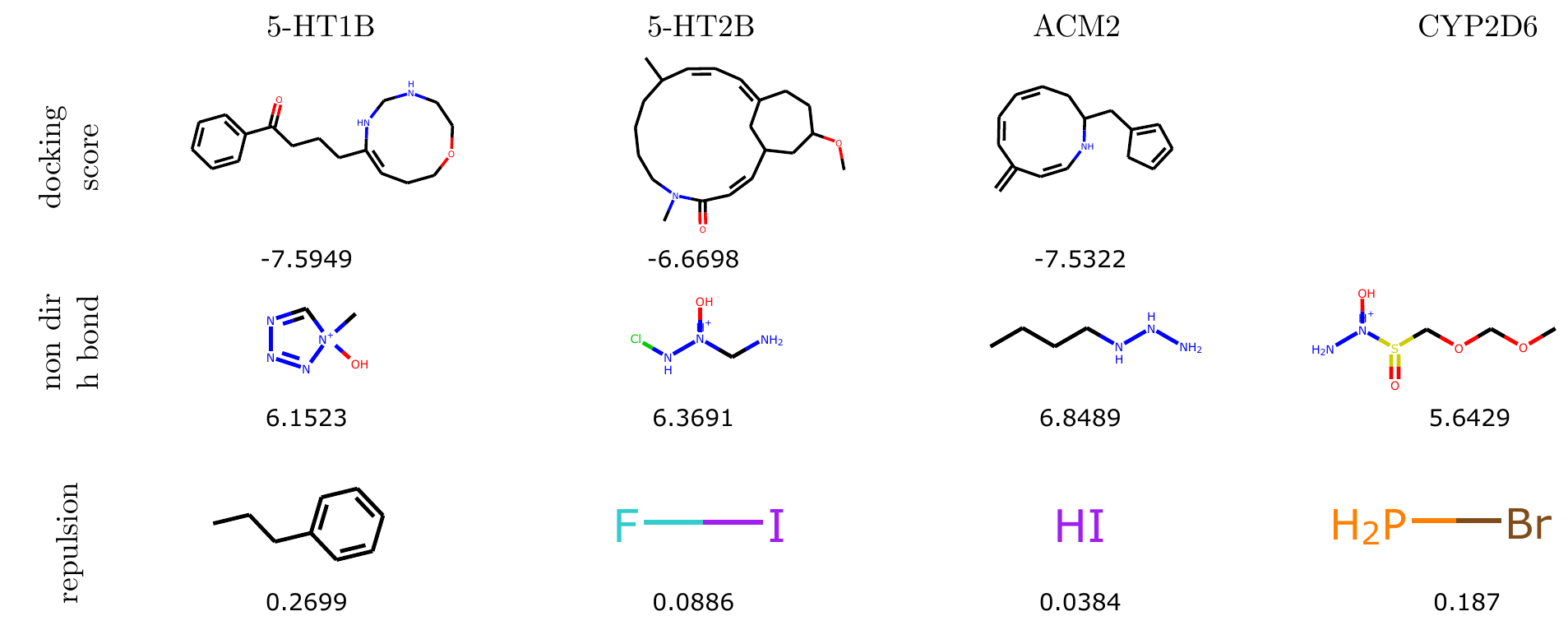}
    \caption{Best scoring molecules generated by CVAE in each of the three tasks composing the benchmark. Missing compounds correspond to the failed optimizations.}
    \label{fig:benchmark_visualization}
\end{figure}

\begin{figure}
    \centering
    \includegraphics[width=\textwidth]{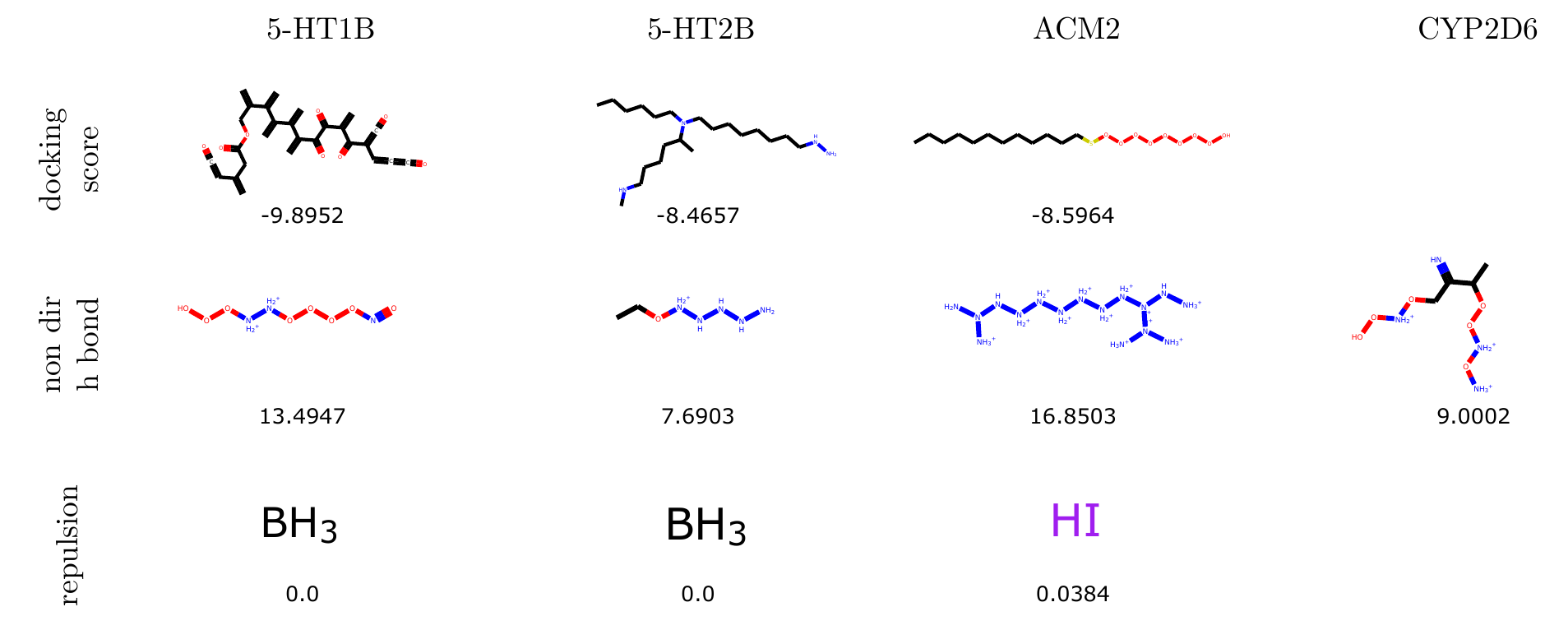}
    \caption{Best scoring molecules generated by GVAE in each of the three tasks composing the benchmark. Missing compounds correspond to the failed optimizations.}
    \label{fig:benchmark_visualization}
\end{figure}

\paragraph{Generated molecules} Figure~\ref{fig:benchmark_visualization} shows the best scoring molecules generated by REINVENT. We observe that optimizing each objective promotes different structural motifs. For example, the best scoring molecules in the \textsc{Repulsion} task are small, which intuitively enables them to easily fit into the binding pocket, achieving lower repulsion values than the top 1\% molecules in the training set.

Similarly, there are clear patterns visible in the top molecules of CVAE and GVAE. For example, CVAE generates macrocycles in the task of docking score optimization, while GVAE generates long chains with no cycles when optimizing the same objective. These models also create oxygen or nitrogen chains when optimizing \textsc{Hydrogen Bonding}, and very small molecules (often less than 3 heavy atoms) for the \textsc{Repulsion} task.

\begin{figure}
    \centering
     \begin{subfigure}[b]{0.24\textwidth}
         \centering
         \includegraphics[width=\textwidth]{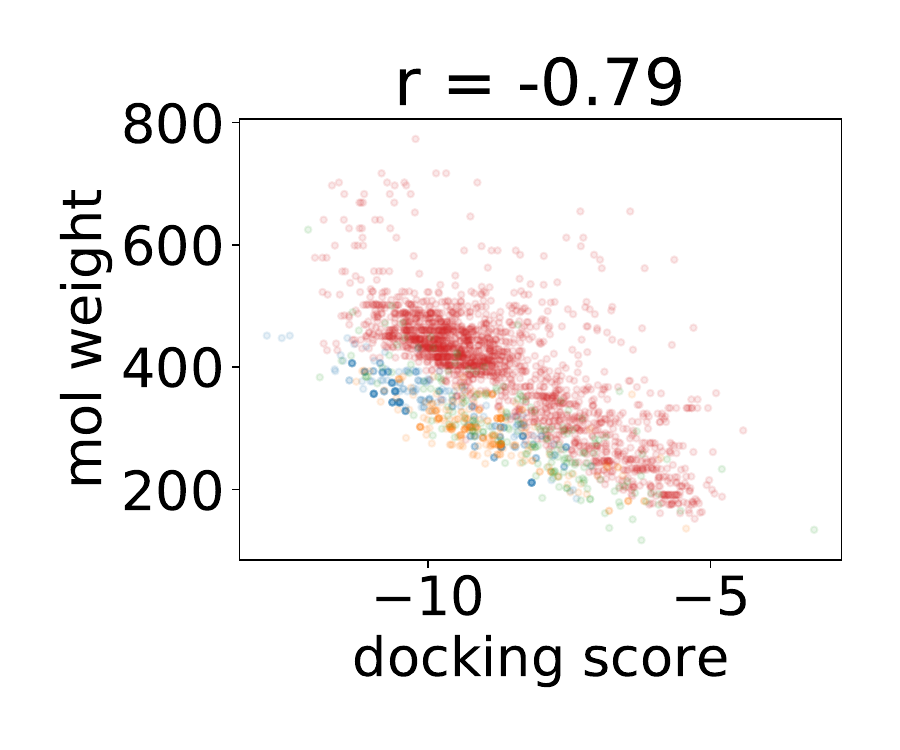}
         \caption{5HT1B}
     \end{subfigure}
     \hfill
     \begin{subfigure}[b]{0.24\textwidth}
         \centering
         \includegraphics[width=\textwidth]{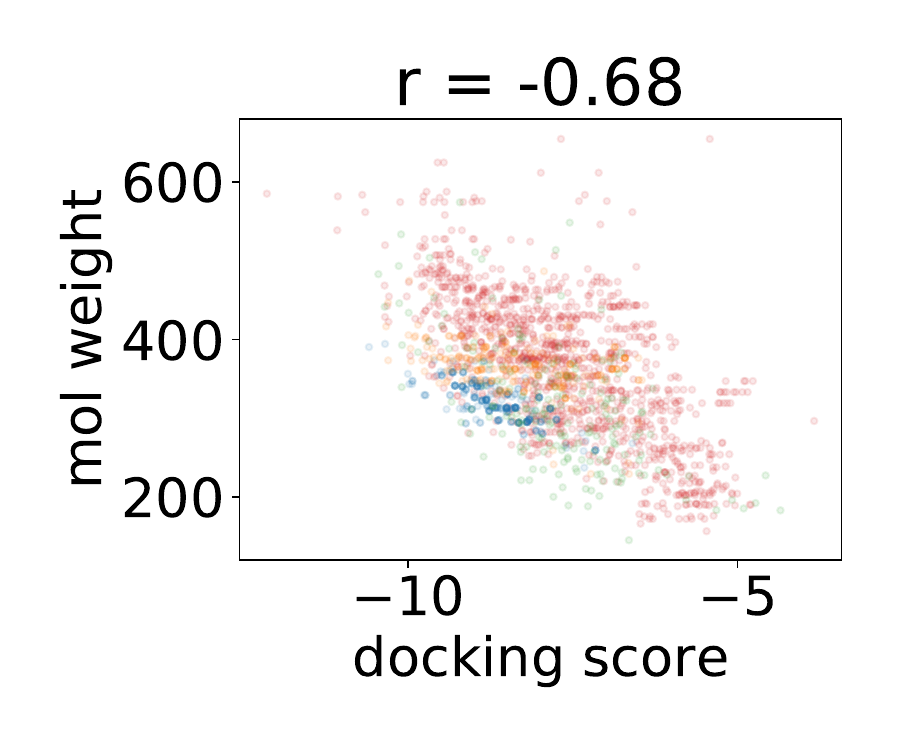}
         \caption{5HT2B}
     \end{subfigure}
     \hfill
     \begin{subfigure}[b]{0.24\textwidth}
         \centering
         \includegraphics[width=\textwidth]{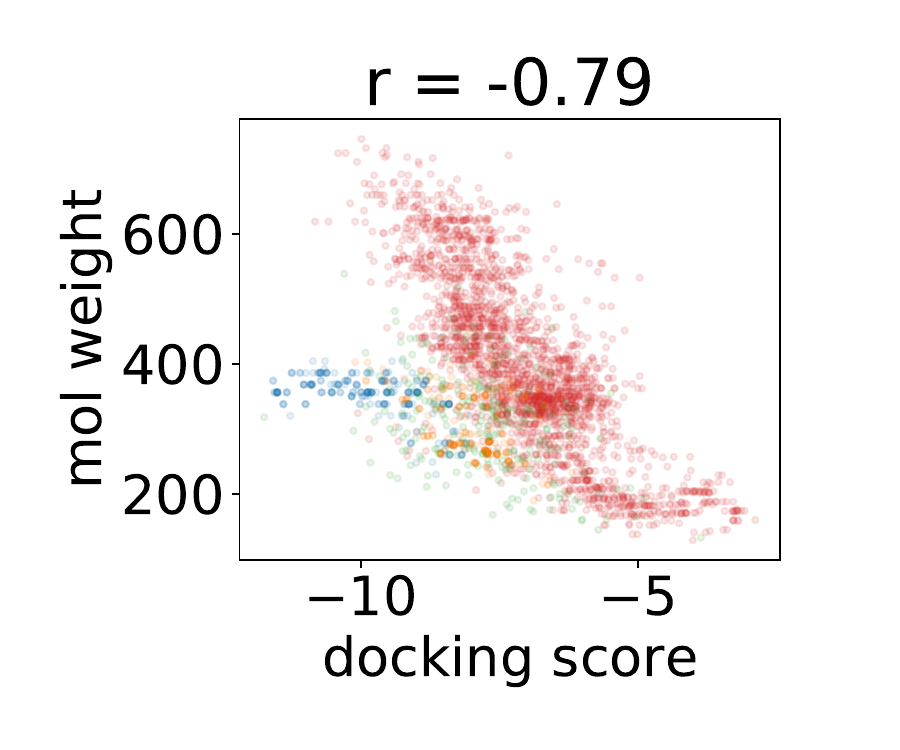}
         \caption{ACM2}
     \end{subfigure}
     \hfill
     \begin{subfigure}[b]{0.24\textwidth}
         \centering
         \includegraphics[width=\textwidth]{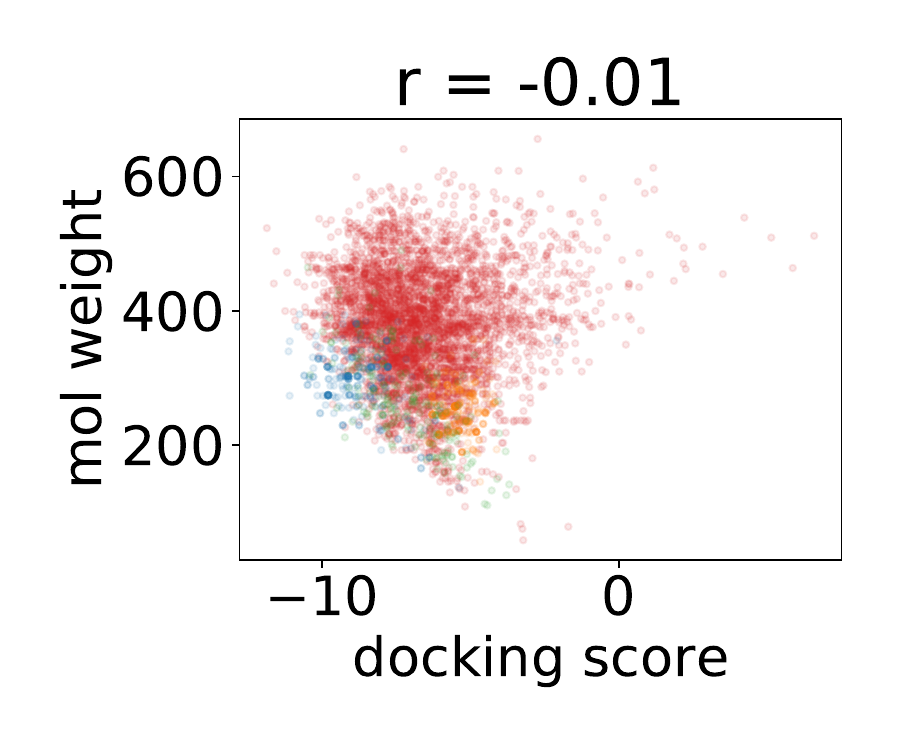}
         \caption{CYP2D6}
     \end{subfigure}
    \caption{Correlation between docking score and molecular weight. The training set is marked with red dots, and the compounds generated by REINVENT by optimizing different targets are colored in blue (\textsc{Docking Score Function}), orange (\textsc{Hydrogen Bonding}), and green (\textsc{Repulsion}).}
    \label{fig:scatter-mw}
\end{figure}

\begin{figure}
    \centering
     \begin{subfigure}[b]{0.24\textwidth}
         \centering
         \includegraphics[width=\textwidth]{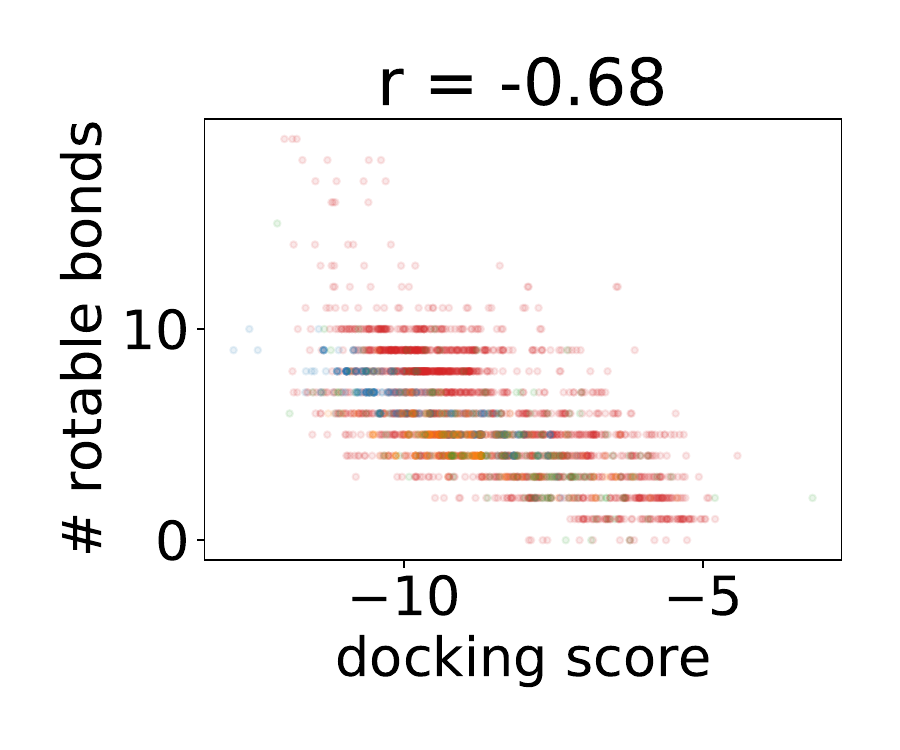}
         \caption{5HT1B}
     \end{subfigure}
     \hfill
     \begin{subfigure}[b]{0.24\textwidth}
         \centering
         \includegraphics[width=\textwidth]{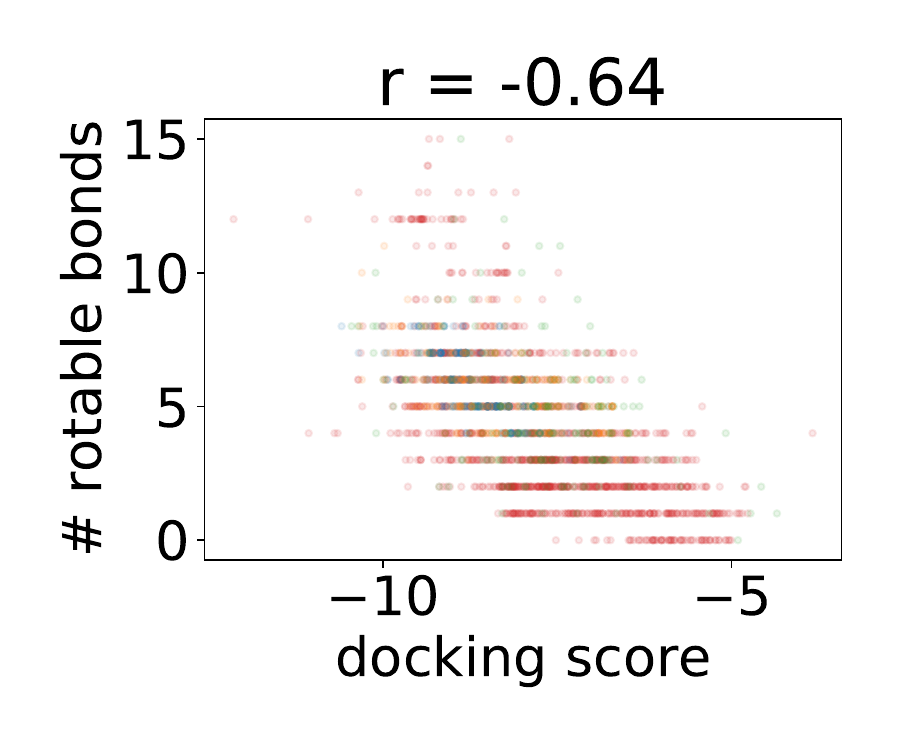}
         \caption{5HT2B}
     \end{subfigure}
     \hfill
     \begin{subfigure}[b]{0.24\textwidth}
         \centering
         \includegraphics[width=\textwidth]{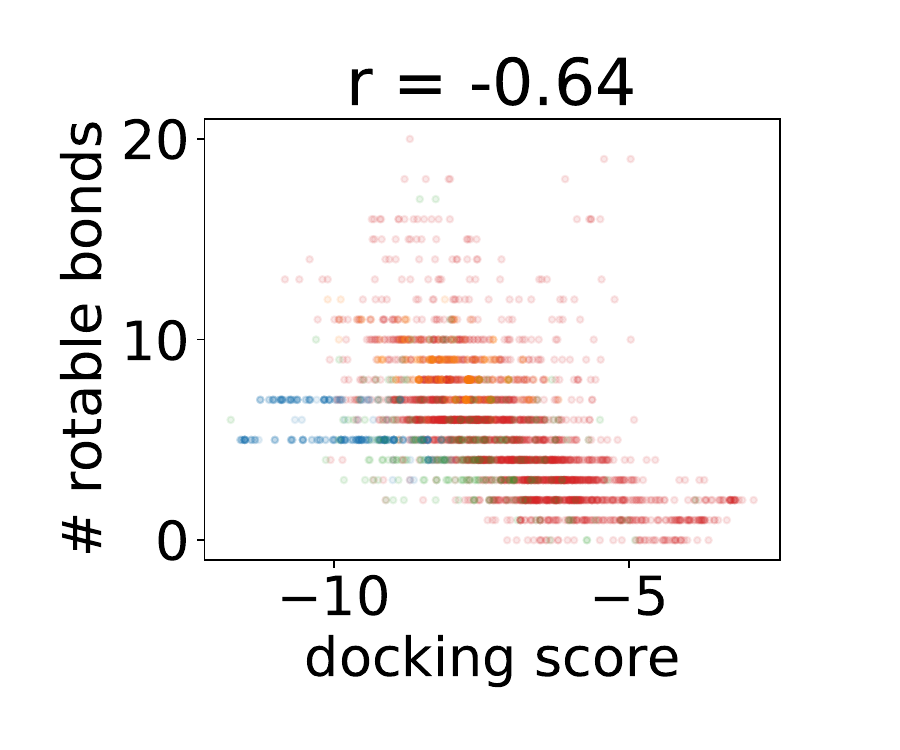}
         \caption{ACM2}
     \end{subfigure}
     \hfill
     \begin{subfigure}[b]{0.24\textwidth}
         \centering
         \includegraphics[width=\textwidth]{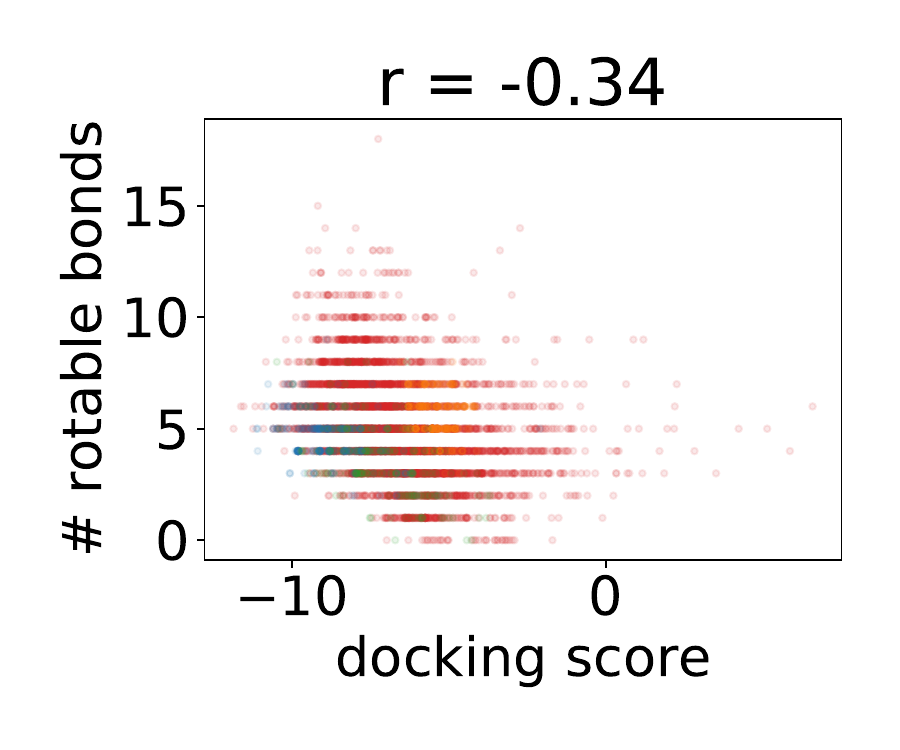}
         \caption{CYP2D6}
     \end{subfigure}
    \caption{Correlation between docking score and the number of rotable bonds. The training set is marked with red dots, and the compounds generated by REINVENT by optimizing different targets are colored in blue (\textsc{Docking Score Function}), orange (\textsc{Hydrogen Bonding}), and green (\textsc{Repulsion}).}
    \label{fig:scatter-rb}
\end{figure}

We noticed a moderately strong correlation between docking scores and the number of rotable bonds or molecular weight. Figures~\ref{fig:scatter-mw} and \ref{fig:scatter-rb} show that with the increasing number of rotable bonds or molecular weight, the docking scores improve. For the number of rotable bonds, the generated compounds are well mixed with the training data marginal distribution. On the other hand, the distribution of generated compounds is shifted towards better docking scores and smaller molecular weights in the case of the weight-to-docking-score relation. In other words, molecules achieve better docking scores at the same molecular weight after the optimization. The correlations are weaker for CYP2D6, which may be caused by a bigger binding site of this enzyme. However, the last observation about molecular weights holds.

When examining generated compounds from the chemical point of view, it should undeniably be stated that REINVENT produced the most consistent ligands with the highest possibility of desired biological activity. When different optimization approaches are considered, the best results were produced during the docking score optimization. Non-dir h-bond optimization produced compounds with sometimes a high number of moieties able to produce a hydrogen bond. In the repulsion task, the produced compounds are correct from the chemical point of view. The drug-likeliness of compounds produced by CVAE and GVAE is lower (although they still meet criteria included in the Lipiski Rule of Five), but they still can be used in the docking benchmark task.

\section{Conclusion}

As concluded by \citet{coley2019}, ``the current evaluations for generative models do not reflect the complexity of real discovery problems''. Motivated by this, we proposed a new, more realistic, benchmark tailored to de novo drug design, using docking score as the target to optimize. Code to evaluate new models is available at \href{https://github.com/cieplinski-tobiasz/smina-docking-benchmark}{https://github.com/cieplinski-tobiasz/smina-docking-benchmark}.

Our results suggest that generative models applied to de novo drug discovery pipelines might require substantial more data to generate realistic compounds than is typically available for training. Despite using over $1000$ compounds for training (between $1074$ and $3780$), the best docking scores generally do not outperform top 10\% docking scores in the ZINC dataset. Docking score is only a simple proxy of the actual binding affinity, and as such it should worry us that it is already challenging to optimize. 

On a more optimistic note, the tested models achieved much better performance on the simplest task in the benchmark, which is to optimize a single term in the SMINA scoring function involving the number of hydrogen bonds to the binding site. This suggests that producing compounds that optimize docking score based on the provided dataset is an attainable, albeit challenging, task. We hope our benchmark better reflects the complexity of real discovery problems and will serve as a stepping stone towards developing better de novo models for drug discovery.

\bibliography{refs}

\appendix
\section{Default SMINA scoring function}
\label{app:smina_scoring_function}

We include the definitions of SMINA's default scoring function components and weights used for calculating docking score in score only mode. $a_1$ and $a_2$ denote atoms, $d(a_1, a_2)$ is the distance between atoms, $d_\text{opt}$ is the sum of their van der Waals radii and $d_\text{diff}(a_1, a_2) = d(a_1, a_2) - d_\text{opt}(a_1, a_2)$. Distance unit is Angstrom ($10^{-10}\text{m}$).

\begin{equation*}
\label{eq:docking_score}
    \begin{split}
    \textnormal{Docking score} =& -0.035579 \cdot \textnormal{gauss}(o=0, w=0.5) \\
     & -0.005156 \cdot \textnormal{gauss}(o=3, w=2) \\
     & +0.840245 \cdot \textnormal{repulsion} \\
     & -0.035069 \cdot \textnormal{hydrophobic} \\
     & -0.587439 \cdot \textnormal{non\_dir\_h\_bond}
    \end{split}
\end{equation*}

\begin{align*}
    \textnormal{gauss}(a_1, a_2) &= \exp\left(-\left(\frac{d_\textnormal{diff}(a_1, a_2) - o}{w}\right)^2\right) \\
    \textnormal{repulsion}(a_1, a_2) &= \begin{cases}
        d_\textnormal{diff}(a_1, a_2)^2, & d_\textnormal{diff}(a_1, a_2) < 0 \\
        0, & \textnormal{otherwise}
    \end{cases} \\
    \textnormal{hydrophobic}(a_1, a_2) &= \begin{cases}
        0, & not\_hydrophobic(a_1) \; \textnormal{or} \; not\_hydrophobic(a_2) \\
        1, & d_\textnormal{diff}(a_1, a_2) < 0.5 \\
        0, & d_\textnormal{diff}(a_1, a_2) \geq 1.5 \\
        1.5 - d_\text{diff}(a_1, a_2), & \textnormal{otherwise}
    \end{cases} \\
    \textnormal{non\_dir\_h\_bond}(a_1, a_2) \space &= \begin{cases}
        0, & (a_1, a_2) \; \textnormal{do not form hydrogen bond} \\
        1, & d_\textnormal{diff}(a_1, a_2) < -0.7 \\
        0, & d_\textnormal{diff}(a_1, a_2) \geq 0 \\
        \frac{d_\text{diff}(a_1, a_2)}{-0.7}, & \textnormal{otherwise}
    \end{cases}
\end{align*}

\section{Model details}
\label{app:hyperparameters}

We include hyperparameters and training settings used in our models. Our code is available at \href{https://github.com/cieplinski-tobiasz/smina-docking-benchmark}{https://github.com/cieplinski-tobiasz/smina-docking-benchmark}.

MLP is used to predict docking score from CVAE or GVAE latent space representation of molecule. It is a simple feed forward neural network with one hidden layer. Hyperparameters of this model are listed in \ref{tab:app:mlp_hparams}.

\begin{table}[h]
\centering
\begin{tabular}{ l c }
    \hline
    & Parameter \\
    \hline
    Training epochs & 50 \\ 
    Layers number & 1 \\ 
    Hidden layer dim & 1000 \\
    Loss function & Mean Squared Error \\
    Optimizer & Adam \\
    Learning rate & 0.001 \\
    \hline
\end{tabular}
\caption{MLP hyperparameters}
\label{tab:app:mlp_hparams}
\end{table}

Both Chemical VAE and Grammar VAE are based on variational autoencoder model with stacked convolution layers in its encoder part and stacked GRU layers in decoder part. What differs them is the way that SMILES is encoded to one hot vector. Chemical VAE encodes each character of SMILES to separate one-hot vector, while Grammar VAE forms a parse tree from SMILES and encodes the parse rules. Details for CVAE are listed in \ref{tab:app:cvae_hparams} and for GVAE in \ref{tab:app:gvae_hparams}.

\begin{table}[h]
\centering
\begin{tabular}{ l c }
    \hline
    & Parameter \\
    \hline
    MLP learning rate & 0.05 \\
    MLP descent iterations & 50 \\
    Fine-tuning batch size & 256 \\
    Fine-tuning epochs & 5 \\
    Latent space dim & 196 \\
    Encoder convolution layers number & 4 \\
    Decoder GRU layers number & 4 \\
    \hline
\end{tabular}
\caption{Chemical VAE hyperparameters}
\label{tab:app:cvae_hparams}
\end{table}

\begin{table}[h]
\centering
\begin{tabular}{ l c }
    \hline
    & Parameter \\
    \hline
    MLP learning rate & 0.01 \\
    MLP descent iterations & 50 \\
    Fine-tuning batch size & 256 \\
    Fine-tuning epochs & 5 \\
    Latent space dim & 56 \\
    Encoder convolution layers number & 3 \\
    Decoder GRU layers number & 3 \\
    \hline
\end{tabular}
\caption{Grammar VAE hyperparameters}
\label{tab:app:gvae_hparams}
\end{table}

The REINVENT model was used with the default parameters of the original implementation provided by~\citet{olivecrona2017molecular}. The model consists of 3 GRU layers and was pretrained on the ChEMBL dataset. The RL agent was trained for 200 steps with a random forest scoring function. The hyperparameters of the random forest were chosen by a grid search, and the resulting configuration is shown in Table~\ref{tab:app:rf_hparams}.

\begin{table}[h]
\centering
\begin{tabular}{ l c }
    \hline
    & Parameter \\
    \hline
    number of estimators & 500 \\ 
    criterion & gini \\
    maximum number of features in splits & sqrt(number of features) \\
    \hline
\end{tabular}
\caption{RF hyperparameters}
\label{tab:app:rf_hparams}
\end{table}

\section{Dataset details}
\label{app:dataset_details}

The compound sets were downloaded from the ChEMBL database~\citep{chembl2016}. All records referring to human- and rat-based records were taken into account. Compounds with Ki values below 100 nM (referred to as active compounds) and above 1000 nM (inactive ones) were taken into account. Only binding data were considered, and it was assumed that IC50 = Ki/2. The compound protonation states were generated for pH = 7.4.

The crystal structures for docking were fetched from the PDB database, the following structures were used in the study: 4IAQ for 5-HT1B, 4NC3 for 4-HT2B, 3UON for ACM2, and 3QM4 for CYP3D6. The Protein Preparation Wizard from the Schrödinger molecular modeling package was used for protein preparation for docking. 
\end{document}